\documentclass[12pt,preprint]{aastex}

\shorttitle{A NIR STUDY OF THE NGC 7538 STAR-FORMING REGION}

\shortauthors{OJHA ET AL.}

\begin{document}

\title{A near-infrared study of the NGC 7538 star forming region}

\author{D.K. Ojha\altaffilmark{1}, M. Tamura, Y. Nakajima, and M. Fukagawa}
\affil{National Astronomical Observatory of Japan, Mitaka, Tokyo 181-8588, Japan; 
ojha@tifr.res.in}
\and
\author{K. Sugitani}
\affil{Institute of Natural Sciences, Nagoya City University, Mizuho-ku, 
Nagoya 467-8501, Japan}
\and
\author{C. Nagashima, T. Nagayama, T. Nagata, and S. Sato}
\affil{Department of Astrophysics, Faculty of Sciences, 
Nagoya University, Chikusa, Nagoya 464-8602, Japan}
\and
\author{S. Vig, and S.K. Ghosh}
\affil{Tata Institute of Fundamental Research, Homi Bhabha Road, Colaba,
Mumbai - 400 005, India}
\and
\author{A.J. Pickles}
\affil{Caltech Optical Observatories, Caltech 105-24 Astronomy, 
Pasadena CA 91125, USA}
\and
\author{M. Momose}
\affil{Institute of Astrophysics and Planetary Sciences, 
Ibaraki University, Bunkyo 2-1-1, Mito, Ibaraki 310-8512, Japan}
\and 
\author{K. Ogura}
\affil{Kokugakuin University, Higashi, Shibuya-ku, Tokyo 150-8440, Japan}

\altaffiltext{1}{On leave from Tata Institute of Fundamental Research, 
Mumbai (Bombay) - 400 005, India (ojha@tifr.res.in)}

\begin{abstract}

We present sub-arcsecond (FWHM $\sim$ 0\arcsec.7), near-infrared (NIR) 
JHK$_s$-band images and a high sensitivity radio continuum image at
1280 MHz, using \mbox{SIRIUS} on UH 88-inch telescope and GMRT. 
The NIR survey covers an area of 
$\sim$ 24 arcmin$^2$ with 10 $\sigma$ limiting magnitudes of $\sim$ 19.5, 
18.4, and 17.3 in  J, H, and K$_s$-band, respectively. 
Our NIR images are deeper than any JHK surveys 
to date for the larger area of NGC 7538 star forming region. We construct JHK 
color-color and J-H/J and \mbox{H-K/K} color-magnitude diagrams to identify 
young stellar objects (YSOs) and to estimate their masses. Based on these 
color-color 
and color-magnitude diagrams, we identified a rich population of YSOs 
(Class I and Class II), associated with the NGC 7538 region. A large number of 
red sources \mbox{(H-K $>$ 2)} have also been detected around NGC 7538. We 
argue that these red stars are most probably pre-main sequence stars with 
intrinsic color excesses. Most of YSOs in NGC 7538 are arranged 
from the north-west toward south-east regions, forming a sequence in age:
the diffuse H II region (north-west, oldest: where most of the Class II and 
Class I sources are detected); the compact IR core (center); and the 
regions with the extensive IR reflection 
nebula and a cluster of red young stars (south-east and south). 
We find that the slope of the K$_s$-band luminosity 
function of NGC 7538 is lower than the typical values reported for the young 
embedded clusters, although equally low values have also been reported in the
W3 Main star forming region. From the slope of the K$_s$-band luminosity 
function and the analysis by Megeath et al. (1996), we infer that the
embedded stellar population is comprised of YSOs with an age of $\sim$ 1 Myr. 
Based on 
the comparison between models of pre-main sequence stars with the observed 
color-magnitude diagram we find that the stellar population in NGC 7538 is 
primarily composed of low mass pre-main sequence stars similar to those 
observed in the W3 Main star forming region. 
The radio continuum image from the GMRT observations at 1280 MHz shows an
arc-shaped structure due to the interaction between the
H II region and the adjacent molecular cloud. The ionization front at the
interface between the H II region and the molecular cloud is clearly 
delineated by
comparing the radio continuum, molecular line, and near-infrared images.
\end{abstract}


\section{Introduction} 

The H II region NGC 7538, which is part of the Cas OB2 complex at a distance 
of 2.8 kpc (Blitz, Fich, \& Stark 1982; Campbell \& Thompson 1984), 
harbors massive stellar objects in various early 
stages of development (Campbell \& Persson 1988, and references therein). 
The NGC 7538 star forming region consists of an
optically visible H II region and at least 11 high luminosity infrared
sources (NGC 7538 IRS 1--11), probably corresponding to young massive
stars (Kameya et al. 1990). There are three major activity centers in
\mbox{NGC 7538}; the north-west region contains IRS 4--8, the central region 
IRS 1--3, and the south-east companion IRS 9 (Werner et al. 1979).
McCaughrean, Rayner, \& Zinnecker (1991) discussed their 
NIR images from a morphological perspective and suggested several different 
evolutionary stages in the NGC 7538 star forming complex with considerable
substructure. The central region hosts a pair of ultra-compact
(UC) H II regions, NGC 7538A and NGC 7538B (Wood \& Churchwell 1989). 
The far-infrared luminosity from IRS 1--3 in the central region 
is 2.5$\times$10$^5$ L$_{\odot}$
(Werner et al. 1979). The luminosity of IRS 9 is 4$\times$10$^4$ L$_{\odot}$.  
NGC 7538 IRS 1 is associated with a large variety
of molecular maser species (Dickel et al. 1982; Kameya et al. 1990;
Menten et al. 1986; Rots et al. 1981; Madden et al. 1986; Gaume et al. 1991).
VLA radio observations 
were made by Rots et al. (1981) and Campbell (1984). Campbell (1984)
presented strong evidence for the existence of a collimating disk around 
NGC 7538 IRS 1 and an associated outflow of ionized gas. 

Momose et al. (2001) presented imaging polarimetry of the 850 $\mu$m 
dust continuum emission in the NGC 7538B region with the SCUBA 
Polarimeter on JCMT. 
They found two prominent cores associated with IRS 1 and IRS 11 
in the surface brightness map of the continuum emission. The total cloud mass
derived from the surface brightness map is 6.7$\times$10$^3$ M$_{\odot}$.
The polarization map shows a striking difference between IRS 1 and 
IRS 11, suggesting that small scale fluctuations of the magnetic field are
more prominent in IRS 1. They interpreted this in terms of a difference
in evolutionary stage of the cores. Inside IRS 1, which seems to be at a later 
evolutionary stage than IRS 11, substructures such as subclumps or a cluster
of infrared sources have already formed. A 2 mm continuum mapping observation
around NGC 7538 IRS 1--3 was made by Akabane et al. (2001) using the 
Nobeyama Bolometer Array (NOBA) mounted on the Nobeyama 45m telescope.
They derived the total cloud mass of about 6.9$\times$10$^3$ M$_{\odot}$,
which is almost the same as that of Momose et al. (2001).

Bloomer et al. (1998) imaged the NGC 7538 IRS 1--3 region in various
infrared wavelength bands. They studied in detail the infrared-bright knots 
due to  
very young O-type stars (IRS 1, 2, and 3) behind an extinction of at
least A$_V$ = 16 mag. 
They found a shell-like structure which presumably corresponds to a shock
caused by the high-velocity stellar wind from IRS 2 colliding with the 
surrounding molecular cloud.
The K$_s$-bright source, IRS 9, was discovered at the tip of the south-eastern
nebula and was identified as a protostellar object by Werner et al . (1979).
There are powerful bipolar outflows from the three dominant 
IR sources in the region, IRS 1, IRS 9 and IRS 11 (Kameya et al. 1989; 
Davis et al. 1998) of which IRS 1 and IRS 11 are well-known OH maser sources. 
IRS 1 and IRS 9 are identified as the most likely powering sources
of massive, bipolar CO outflows, based on the morphology and polarization
of their associated near-infrared reflection nebulae (Tamura et al. 1991).
Numerous H$_2$ features have been observed in the NGC 7538 region, 
in particular a collimated ``jet'' associated with the IRS 9 outflow, as
well as possible bow shocks in both the IRS 1 and IRS 9 flows 
(Davis et al. 1998). These are closely associated with the bipolar CO 
outflows.

The high spatial resolution submillimeter maps show filamentary dust ridges 
extending from and connecting to the three major activity centers (IRS 1--3, 
IRS 11, and IRS 9) in the NGC 7538 star forming cloud 
(Sandell \& Sievers 2004). In addition to the three already known star 
formation centers, Sandell \& Sievers (2004) find a fainter
extended submillimeter source near IRS 4 inside the 
optical H II region. Sandell et al. (2003) reported the detection of 
a massive rotating disk 
around the high mass Class 0 candidate NGC 7538 S both in continuum and in 
molecular lines, about 80\arcsec~to the south of IRS 1. NGC 7538 S is the 
strongest submillimeter continuum source in NGC 7538 
(Sandell \& Sievers 2004). 

In this paper we present high resolution J, H, and K$_s$-band
observations over a large area ($\sim$ 24 arcmin$^2$) centered on 
IRS 1--3. To complement NIR data and to understand the H II region more
comprehensively radio continuum observations are also obtained. 
As in our previous work on the W3 Main star forming region 
(Ojha et al. 2004), our motivation is to look for new 
YSOs associated with the NGC 7538 region and to classify their evolutionary 
stages. From the slope of the K$_s$-band luminosity function (KLF), we 
discuss the age-sequence 
and the mass-spectrum of YSOs in the NGC 7538 region. 
In the case of W3 Main we find that the power-law slope of the 
KLF is lower than the typical values 
reported for embedded young clusters. Another motivation of this work is 
to compare the KLF slope of NGC 7538 with that of W3 Main to find any 
similarity or difference in the properties of the two high-mass star forming 
regions and of the sub-regions within the NGC 7538.
 
In \S 2 we present the details of observations and data 
reduction procedures, \S 3 deals with the results and discussion on
mostly point-like YSOs and describe the details of individual nebulosity
and YSO. We then summarize our conclusions in \S 4. 

\section{Observations and data reduction}

\subsection{NIR observations}

The sub-arcsec imaging observations of the NGC 7538 star forming region 
in the NIR wavelengths J ($\lambda$ = 1.25 $\mu$m), 
H ($\lambda$ = 1.63 $\mu$m), and K$_s$ ($\lambda$ = 2.14 $\mu$m) were obtained 
on 2000 August 15 with the University of Hawaii 2.2 m telescope and SIRIUS 
(Simultaneous three-color InfraRed Imager for Unbiased Surveys), 
equipped with three 1024$\times$1024 HgCdTe arrays. The field of view in each 
band is $\sim$ 4\arcmin.9 $\times$ 4\arcmin.9, with a pixel scale of 
0\arcsec.28 at the $f/10$ Cassegrain focus. The HgCdTe arrays work linearly 
within 3\% upto 15,000 ADU and saturate at $\sim$ 25,000 ADU 
(Nagayama et al. 2003). At K$_s$ = 12 mag, the ADU counts are well below 
15,000, thus we consider the source magnitudes to be correct within 3\%. 
Further details of the instrument are given in Nagashima et al. (1999) and
Nagayama et al. (2003). 

We obtained 27 dithered exposures of the target centered at 
($\alpha$, $\delta$)$_{2000}$ = ($23^h13^m43^s.95$, 
+61$^{\circ}28^{\arcmin}44^{\arcsec}.2$), 
each 20s long, simultaneously for each band and 9 dithered sky frames
centered at ($\alpha$, $\delta$)$_{2000}$ = ($23^h15^m59^s.60$, 
+61$^{\circ}28^{\arcmin}44^{\arcsec}.2$),  
which is $\sim$ 34\arcmin~east of the target position. The sky frame 
was also used as a reference field 
to assess the stellar populations within the NGC 7538 star forming 
region (see \S 3). Total on-target
integration time in each of the bands was 9 minutes. All the observations 
were done
under good photometric sky conditions. We found an rms magnitude 
fluctuation of less than 0.04 mag in JHK$_s$-bands during 
the NGC 7538 observations. The average seeing size (FWHM) in 
the J, H, and K$_s$-band was 0\arcsec.7, 0\arcsec.7, and 0\arcsec.6, 
respectively during the observations. The observations were
made at air masses between 1.3 and 1.5. Dark frames and dome flats were
obtained at the beginning and end of the observations. The photometric
calibration was obtained by observing the standard star 9183 
in the faint NIR standard star catalog of Persson et al. (1998)       
at air masses closest to the target observations.
The J, H, and K$_s$-band images of the NGC 7538 star forming region are shown 
in Fig. 1. 

Data reduction was done using the pipeline software based on NOAO's 
IRAF\footnote{IRAF is distributed by the
National Optical Astronomy Observatories, which are operated by the
Association of Universities for Research in Astronomy, Inc., under contract
to the National Science Foundation.} package tasks. Dome flat-fielding and
sky subtraction with a median sky frame were applied. Identification and
photometry of point sources were performed by using the DAOFIND and DAOPHOT
packages in IRAF, respectively. Because of source confusion and nebulosity
within the NGC 7538 region, photometry was obtained using the point-spread 
function (PSF) algorithm ALLSTAR in the DAOPHOT package (Stetson 1987). For
the JHK$_s$-band images the adopted fitting radii were 3 pixels 
($\sim$ 1 FWHM of the PSF), and the PSF radius was 13 pixels. The local sky was
evaluated in an annulus with an inner radius of 12 pixels and a width of
21 pixels. We used an aperture radius of 3 pixels ($\sim$ 0\arcsec.84)
with appropriate aperture corrections per band for the final photometry.   

The resulting photometric data are in the
SIRIUS system. For the purposes of plotting these data in 
color-color and color-magnitude diagrams,
we have converted them into the California Institute of Technology (CIT) system
using the color transformations between the SIRIUS and CIT systems
(Nagashima et al. 2003)\footnote{also available at 
http://www.z.phys.nagoya-u.ac.jp/$\sim$sirius/about/color$_{-}$e.html}, which 
have been obtained by observing several of the red standard stars of
Persson et al. (1998). Absolute position calibration was achieved using the    
coordinates of a number of stars from the 2MASS catalogue. The position
accuracy is better than $\pm$0.\arcsec05 rms in the NGC 7538 field.
 
The completeness limits of the images were evaluated by adding artificial
stars of different magnitudes to the images and by determining the fraction
of these stars recovered in each magnitude bin. The recovery rate was greater
than 90\% for magnitudes brighter than 19, 18.5, and 17.5 in the J, H, and 
K$_s$-bands respectively. The observations are complete (100\%) to the level
of 17, 16 and 15 magnitudes in J, H and K$_s$-bands respectively. The
limiting magnitudes (at 10 $\sigma$) are estimated to be 
$\sim$ 19.5, 18.4, and 17.3 at J, H, and K$_s$-bands, respectively. 
We found that within the 10 $\sigma$ detection limit, the accuracy on 
magnitudes of $\sim$ 98\% stars in our sample is better than 0.2 mag. 
The sources are saturated at K$_s$ $<$ 12. 

We estimated the errors in photometry due to source confusion and
nebulosity through artificial star experiments. The difference between the
magnitudes of the added and recovered stars should reflect the effect of
confusion with other stars and the background nebulosity. 
We find that for J = 19.5,
H = 18.4, and K$_s$ = 17.3 stars (at our 10 $\sigma$ mag detection limit), the
rms error of the difference is 0.17, 0.15, and 0.14 mag, respectively.
The rms error of the difference is 0.14, 0.15, and 0.15 mag, respectively
for J = 19, H = 18.5, and K$_s$ = 17.5 stars (corresponding to the 90\%
completeness level). The error increases rapidly with increasing magnitude.

\subsection{GMRT radio continuum observations}

The ionized gas within and around the H II region associated with NGC 7538
star forming region has been mapped at low and high angular resolutions 
using the Giant Metrewave Radio Telescope (GMRT) array in India. 
The radio continuum 
observations centered at ($\alpha$, $\delta$)$_{2000}$ = 
($23^h13^m43^s.95$, +61$^{\circ}28^{\arcmin}44^{\arcsec}.2$) have been 
carried out at 1280 MHz on 2004 January 26.
The source 3C48 was used as the primary flux calibrator for
1280 MHz observations, while the source 2355+498 was used as a secondary
calibrator. The GMRT antennas and their configurations are discussed in
detail by Swarup et al. (1991).
 
Data reduction was done in classic AIPS. Bad data (dead antennas,
interference, spikes, etc.) were identified and flagged using UVFLG and
TVFLG. Images of the field were formed by Fourier inversion and cleaning
(IMAGR). The initial images were improved by self-calibration (CALIB) in
both phase and amplitude.
 
Fig. 2 shows the radio continuum image of the whole NGC 7538 region, 
generated from the GMRT interferometric observations at 1280 MHz. 
This image has a resolution of 
22\arcsec.5 $\times$12\arcsec.4 and an rms of $\sim$ 4 mJy beam$^{-1}$.
The total flux density is 13.1 Jy in the map. The inset in Fig. 2 is a high
resolution GMRT map of the central radio peak. This image has a
resolution of 6\arcsec.6 $\times$ 4\arcsec.7, an rms of 
$\sim 1.3$ mJy beam$^{-1}$, and the total flux density of 978.7 mJy.
The primary FOV at 1280 MHz is about 0.4 degree. 

\section{Results and Discussion}

\subsection{Prominent infrared nebulae}

A composite color image is constructed from the SIRIUS J, H, and K$_s$-band 
images (J represented in blue, H in green, and K$_s$ in red) and is
shown in Fig. 3. The individual prominent IR sources are marked in the 
K$_s$-band image in Fig. 1. The most prominent features are
the diffuse purple-pink color to the N-W, the compact and bright orange
color at the center and the fluffy and dark orange color to the S-E. It is
noteworthy that the scarce star field (black) extends to the S-W of the
field, where the dense CO molecular cloud is located (see Fig. 4 and 
Davis et al. 1998). We briefly describe the infrared appearance of the
nebulae from a morphological perspective here, then discuss the 
point-like sources in detail below. 

The large diffuse emission extending to N-W of IRS 1, 2, and 3 
(see Fig. 3) is probably due to the combination of free-free 
and bound-free emissions, corresponding to what is seen optically and 
coincides well with the radio brightness of GMRT at 1280 MHz shown in Fig. 4.
The extended filaments of H$_2$ line emission to the N-W of the IRS 1--3
cluster around the photo-dissociation regions (PDRs) at the interface
between H II and molecular cloud are particularly striking (Fig. 5 and 
Davis et al. 1998). 

The bright and compact infrared nebula embedded with IRS 1, 2, and 3 at the
center of Fig. 3 (see also Fig. 12a) is coincident with the peak of radio
continuum, although the IRS sources remain unresolved with GMRT.
The radio peak in the center of the map in Fig. 2
appears elongated northward by 60\arcsec. The high resolution radio map 
(inset in Fig. 2) resolves this elongation into two lobes with a separation of 
26\arcsec~in the N-S direction.

The fluffy nebula associated with IRS 9 is revealed 
$\sim$ 2\arcmin~south-east of IRS 1. A few red young stars are 
located at the easternmost tip of the nebula.
The fluffy morphology as well as the dark patches extending 70\arcsec~to 
the west plausibly arises from reflection illuminated by these red
sources (see Fig. 12b and \S 3.8.2). The direction of the NGC 7538 IRS 9
``H$_2$ jet'' reported by Davis et al. (1998) is indicated in Fig. 5.
Two small peculiar nebulosities are also revealed at the north-east corner in 
the composite image of Fig. 3 and in the continuum-subtracted H$_2$ image of 
Fig. 5. The enlarged view of this region is shown in Fig. 12d.  

The selected regions with individual IR sources and associated nebulosities 
are described in detail in \S 3.8.

\subsection{Ionizing sources of H II region}

In the diffuse H II region, there lie five radio peaks in Fig. 4. Although
three bright IRS sources lie therein, none of them coincide with any radio 
peaks. IRS 5, located close ($\sim$ 6\arcsec~N-W) to the brightest 
radio peak and assigned as O9 type star in the H-K vs. K color-magnitude 
diagram (see Fig. 7; Moreno \& Chavarria 1986), might be responsible for the 
nearby radio peak, but is still insufficient to power the whole H II region. 
IRS 6, the brightest infrared source in the H II region, is located 
$\sim$ 27\arcsec~($\sim$ 0.37 pc in projection) north-east of the brightest 
radio peak and is estimated to be an O6-O7 type from its color and IR 
luminosity (Wynn-Williams et al. 1974; Moreno \& Chavarria 1986). 
IRS 7 has been identified as a K type foreground star 
(Moreno \& Chavarria 1986), and IRS 8 does not coincide with any optical
or radio feature (Wynn-Williams et al. 1974). IRS 7 and IRS 8 are most 
probably not associated with the NGC 7538 nebula (Wynn-Williams et al. 1974;
Zheng et al. 2001). 

It is most plausible that IRS 6 is the main exciting source responsible for 
this H II region. However, the theoretical Stromgren radius
of the H II region ionized by an O6 type star is roughly fifteen times
larger than the observed one (see Fig. 3). 
It means that the H II region is under expansion and the central star (IRS 6)
is at a very early stage of its evolution. But the number of UV photons
emitted by O stars dramatically changes with the spectral sub-class. In
addition, young O stars are often surrounded by some amount of dust, which
attenuates the UV emission to ionize the surrounding gas. As a result the
size of the Stromgren sphere is reduced compared to the ideal case.
Therefore, the precise spectral type of the exciting star and information on
the amount of the surrounding dust are required for a quantitative
discussion of the evolution of the NGC 7538 H II region.

In Figs. 4 \& 5, a cavity wall is apparent with a steep cliff running from
N-W to S-E of both CO and radio emission (but with opposite gradients
in the N-E/S-W direction), and a gradual decrease of the radio emission
as well as the H$\alpha$ emission (see Moreno \& Chavarria 1986) from
S-W to N-E. At the cliff the H II region is plausibly ionization-bounded,
while it is density-bounded toward the N-E.


\subsection{Photometric analysis of point-like sources}

\subsubsection{Color-Color diagram}

We obtained photometric data of 1364 sources in J, 1679 in H, and 1682 in
K$_s$-band. Figs. 6a and 6b show the J-H/H-K color-color (CC) diagrams of the 
NGC 7538 star forming region and the reference field, respectively, for the 
sources detected in the JHK$_s$ bands with a positional agreement of less than 
2\arcsec~and with photometric errors in each color of less than 0.1 mag. 
The reference field is also used for the correction of field star 
contamination in the raw K$_s$-band luminosity function of NGC 7538 
(see \S 3.5). In Figs. 6a and 6b, the solid
and broken heavy curves represent the unreddened main sequence and
giant branches (Bessell \& Brett 1988) and the parallel dashed lines are the
reddening vectors for early and late type stars (drawn from the base and tip
of the two branches). The dotted lines indicate the locus of T-Tauri stars
(Meyer et al. 1997). We have assumed that A$_J$/A$_V$ = 0.282,
A$_H$/A$_V$ = 0.175, and A$_K$/A$_V$ = 0.112 (Rieke \& Lebofsky 1985).
As can be seen in Fig. 6a, the stars in NGC 7538 are distributed
in a much wider range than those in the reference field (Fig. 6b),
which indicates that a large fraction of the observed sources in NGC 7538
exhibit NIR excess emission characteristics of young stars with circumstellar
materials as well as a wide range of reddening. 
We classified the sources into three regions in 
the CC diagram (see e.g. Tamura et al. 1998; Sugitani et al. 2002; 
Ojha et al. 2004). ``F'' sources are located between reddening vectors 
projected from the intrinsic color of main-sequence stars and giants and are 
considered to be field stars (main-sequence stars, giants), or 
Class III / Class II sources having small near-infrared excess.
``T'' sources are located redward of region ``F'' but blueward of the reddening
line projected from the red end of the T-Tauri locus of 
Meyer et al. (1997). 
These sources are considered to be mostly classical T-Tauri
stars (Class II objects) with large near-infrared excess.
There may be an overlap in NIR colors of Herbig Ae/Be stars and T-Tauri stars
in the ``T'' region (Hillenbrand et al. 1992). ``P'' sources
are those located in the region redward of region ``T'' and are most likely
Class I objects (protostar-like objects). 
The total number of ``T'' (Class II) and ``P''
(Class I) sources are 268 and 18, respectively.  
By de-reddening the stars on the CC diagram that fell within the reddening
vectors encompassing the main sequence and giant stars, we found the amount of 
visual extinction (A$_V$) for each star. The individual extinction values
range from 0 to 40 magnitudes with an average extinction of 
A$_V$ $\sim$ 7 mag. 

\subsubsection{Color-Magnitude diagram}

The color-magnitude (CM) diagram is a useful tool to study the nature of the
stellar population within star forming regions and also to estimate
its spectral types. Fig. 7 is the H-K vs K CM diagram where all
the sources detected in JHK$_s$ bands, plus some 180 stars fainter than our
limit at J-band but still above the detection threshold in H and K$_s$-bands
are plotted. 
An apparent main sequence track is noticeable with A$_V$ $\sim$ 4 mag in this
diagram; but a comparison of it with a similar diagram for the stars in
the reference field shows that it is a false sequence caused by field stars
in the foreground of the NGC 7538 region.
The vertical solid lines (from left to right in Fig. 7) represent 
the main sequence curve reddened by A$_V$ = 0, 20, 40 and 60 mag,
respectively. We have assumed a distance of 2.8 kpc to the sources to reproduce
the main sequence data on this plot. The parallel slanting lines in Fig. 7
trace the reddening zones for each spectral type. YSOs (Class II and I) 
found from the CC diagram (Fig. 6a) are shown as stars and filled triangles,
respectively. 
However, it is important to note that even those sources not shown with stars 
or 
filled triangles may also be YSOs with an intrinsic color excess, since some
of them are detected in the H and K$_s$-bands only and are not in the J-band 
due to their very red colors. Three bright and very red objects 
(K $<$ 12.0, H-K $>$ 4.0) located in the upper right corner of the figure are 
the very young stars IRS 1, IRS 9, and IRS 9N1 in their earliest evolutionary 
phases (see \S 3.8 and Figs. 12a and 12b).
The bright infrared sources labeled as IRS numbers in Fig. 7 are 
associated with the H II region and the molecular cloud (see Figs. 1, 4 \& 5). 
These sources are shown in Table 1.  

\subsection{Spatial distribution of YSOs and cool red sources}

In our deep NIR observations 46 very red sources are detected only in 
the H, and K$_s$-bands. These sources have colors redder than H-K $>$ 2
in Fig. 7. In Fig. 8, the spatial distribution 
of the YSO candidate sources identified in Figs. 6a and 7 are shown. Stars 
(in blue color) represent sources of T-Tauri type (Class II), filled triangles 
(in green color) indicate protostar-like objects (Class I), and 
filled circles (in red color) denote the very red sources (H-K $>$ 2).

Most of the YSOs in the NGC 7538 are arranged in the north-west and 
south-east regions. Class II or T-Tauri type objects and Class I or
protostar-like objects (located in ``T'' and 
``P'' regions in Fig. 6a, respectively) are distributed in the north-west and
south-east regions over the field, but there is an apparent concentration 
of these sources mainly towards north-west (the optically visible 
H II region). 
Class I type sources are distributed along the interface between the optical
\mbox{H II} region and the molecular cloud region to the west.
Stars with large color indices (H-K $>$ 2) are seen near the dense parts of 
the molecular cloud towards south and south-east (Fig. 4 and Sunada et al., 
in preparation). Most of them are clustered near the 
massive molecular clumps surrounding the luminous infrared sources 
IRS 1--3. Some of them are expected to be members of the embedded stellar
clusters around IRS 1--3 and to the south of IRS 1--3 (i.e. around IRS 11). 
Therefore, these sources associated with the molecular clumps around IRS 9,
IRS 11 and IRS 1--3 (Sunada et al., in preparation) are embedded PMS stars, 
presumably. A few red sources
are also seen around the ionization front at the interface between the 
H II region and the molecular cloud, which might have formed due to the
triggered star formation.        

The average extinction through the molecular cloud in NGC 7538 that hosts
the very red sources is A$_V$ $\sim$ 15 mag (H-K $\sim$ 1). If we assume that 
the large H-K ($>$ 2) color results merely from interstellar reddening 
affecting normal stars, then the extinction value might even exceed 40 mag in 
the molecular cloud where most of the red stars are found. However, with such 
a large amount of absorption, diffuse emissions are unlikely to be detected 
in the NIR. Since most of the red sources are associated with faint diffuse 
emission around IRS 1--3, IRS 9, and IRS 11, this provides evidence that 
these sources are YSOs with intrinsic NIR excess and possibly local extinction 
also. In Fig. 7, a large fraction ($\sim$ 96\%) of these sources are located 
above the straight line drawn from an A0 star parallel to the extinction 
vector. This suggests that they are intermediate mass stars with 
circumstellar materials.

Based on the clustering of different YSO candidates in NGC 7538, 
the overall morphology is of three major condensations, which may form a 
sequence in age : the diffuse H II region (north-west : oldest);
the compact IR core (center); and the regions with the extensive IR reflection 
nebula and a cluster of red young stars (south-east and south : youngest). 
This subject is further discussed in \S 3.7.

\subsection{The K$_s$-band Luminosity Function}

We use the K$_s$-band luminosity function (KLF) to constrain the
initial mass function (IMF) and age of the embedded stellar population in
NGC 7538. To derive the KLF, we have determined the completeness of the data 
through artificial star experiments using ADDSTAR in IRAF. This was
performed by adding fake stars in random positions into the images at 
0.5 magnitude intervals and then by checking how many of the added stars 
could be 
recovered at various magnitude intervals. We repeated this procedure 
at least 8 times. We thus obtained the detection rate as a function of 
magnitude, which is defined as the ratio of the number of recovered 
artificial stars over the number of added stars. 

In order to estimate the foreground and background contaminations, we made use
of both the galactic model by Robin et al. (2003) and
the reference field star counts. The star counts were
predicted using the Besan\c con model of stellar population synthesis
(Robin et al. 2003) 
in the direction of the reference field close to the NGC 7538
(see \S 2), that is also corrected for the photometric completeness.
By comparing the resultant model KLF with that of the
reference field (see Fig. 9), we found that the simulation fits fairly 
well with the data. An advantage of using the model is that we can separate the
contamination from the foreground (d $<$ 2.8 kpc) and the background
(d $>$ 2.8 kpc) field stars. As we saw in \S 3.4, the average
extinction to the NGC 7538 region is A$_V$ $\sim$ 15 mag (H-K $\sim$ 1). Assuming
spherical geometry, then background stars are seen through A$_V$ $\sim$ 30 mag
(2$\times$15 mag).
Therefore, we applied an extinction value of A$_V$ = 30 mag 
(or A$_K$ = 3.36 mag) in simulating the background stars. 
We combined the foreground and background stars to make a whole set of
the contamination field and obtained the fraction of the contaminating stars
over the total model counts. 
Then we scaled the model prediction to the star counts in the reference
field, and subtracted the combined foreground (d $<$ 2.8 kpc) and
background (d $>$ 2.8 kpc with A$_K$ = 3.36 mag) data from the KLF of the
NGC 7538 region.

After correcting for the foreground and background star contamination and 
photometric completeness, the resulting KLFs are presented in Fig. 10 for the 
whole NGC 7538 region. They follow 
power-laws in shape. In Fig. 10, a power-law with a slope $\alpha$ 
[$d N(m_K)/dm_K \propto 10^{\alpha m_K}$, where $N(m_K)$ is the number of
stars brighter than $m_K$]
has been fitted to each KLF (Figs. 10a and 10b) using the linear 
least-squares fitting routine. 
To discuss the relationship between the KLF slopes and the star
forming environment of each region, we divided the NGC 7538 region 
into two rectangular areas : the ``younger'' region 
(IRS 1--3, IRS 9, and IRS 11 regions), and the ``older'' region above 
IRS 1--3 (see Figs. 1 and 8 and \S 3.4).
The derived power-law slopes for the ``younger'', ``older'', and whole 
NGC 7538 regions are shown in Table 2.
We also note that the background subtraction does not significantly change
the power-law slope (Fig. 10 and Table 2).

We find a small increase (1$\sigma$ result, see Table 2) in the KLF slope 
from the ``younger'' region ($\alpha$ $\sim$ 0.27) to the most evolved 
``older'' region ($\alpha$ $\sim$ 0.33) in NGC 7538 (see \S 3.4). This is 
consistent with an expected change in the KLF slope with age if the IMF
is identical in the whole NGC 7538 region (Megeath et al. 1996).   
However, such a small effect is not significant in view of the 
reddening and excess due to circumstellar material. 
We, therefore, assume a coeval population of stars and derive the
KLF of the whole NGC 7538 star forming region for comparison with
other regions and a rough estimate of stellar masses in the next section. 
 
The KLF of the whole NGC 7538 region shows a power-law slope 
that is lower than those generally reported for the young embedded 
clusters ($\alpha \sim 0.4$, e.g. Lada et al. 1991, 1993; Lada \& Lada 2003),
although equally low values have also been reported in the W3 Main
star forming region (Megeath et al. 1996; Ojha et al. 2004). 
Thus, this low value of the slope is
indeed the intrinsic property of the stellar population in this region.
Using the analysis as given in detail by Megeath et al. (1996), the estimated 
KLF slope of the whole NGC 7538 region is roughly consistent with the 
Miller-Scalo IMF if the age of NGC 7538 population is $\sim$ 1 Myr.
 
\subsection{Stellar Mass Estimates}

Fig. 11 shows the CM diagram (J-H vs J) for 286 YSO candidate sources 
identified in Figs. 6a and 7. We estimate the mass of the stellar sources by 
comparing 
them with the evolutionary models of PMS stars (Palla \& Stahler 1999). 
The solid curve in Fig. 11 denotes the loci of 10$^{6}$ yr old PMS stars and the 
dotted 
curve for those of 0.3$\times10^{6}$ yr old PMS stars. Masses range from 0.1
to 4 M$_{\odot}$ from bottom to top, for both curves. 
To estimate the stellar masses, we use the J luminosity rather than that of H or 
K$_s$, as J-band is less affected by the emission from circumstellar
materials (Bertout, Basri, \& Bouvier 1988). 

If the age of NGC 7538 population is $\sim$ 1 Myr as estimated from the KLF
slope (see \S 3.5),
$\sim$ 88\% of the YSO candidates detected in J, H, and 
K$_s$-bands have masses less than 3 M$_{\odot}$, and at least $\sim$ 80\%  
of the stars have masses less than 2 M$_{\odot}$ (Fig. 11). Even if the age 
of the stars is 0.3 Myr, $\sim$ 92\% of the stars have masses below 
3 M$_{\odot}$. 
At the distance of 2.8 kpc, assuming an age of $\sim$ 1 Myr, and an 
extinction at J band of up to $\sim$ 2 mag (A$_V$ $\sim$ 7), the 
J magnitude limit (corresponding to 10 $\sigma$ mag detection limit) 
corresponds to M $\sim$ 0.1 M$_{\odot}$, according to the PMS evolutionary 
tracks from Palla \& Stahler (1999). 
This gives an estimate of 
the lowest mass limits of the detected stars in the NGC 7538 star forming 
region in our sample. 

Therefore, the stellar population in NGC 7538 may be primarily composed of low 
mass PMS stars similar to those observed in the W3 Main star
forming region (Ojha et al. 2004). We also see the presence of lower mass 
stars forming the clusters (e.g. near the IRS 1--3 and IRS 4--8 regions) 
together with the newly formed O-B type stars.
These results 
further support the hypothesis that the formation of high mass stars
is associated with the formation of clusters of low mass stars 
(e.g. Lada \& Lada 1991; Zinnecker, McCaughrean, \& Wilking 1993; 
Persi et al. 1994; Tapia et al. 1997; Ojha et al. 2004). 

\subsection{Star formation activity in NGC 7538}

The NGC 7538 star forming complex seems to be composed of several
structures of different evolutionary stages aligned from north-west to
south-east. The first is the north-western region, which corresponds to the
visible H II region. Here, in addition to the optically visible O-type
stars (e.g., IRS 5 and 6) we detect a large number of Class II and Class I
sources (see Fig. 8). This north-western region is undoubtedly more evolved
than other regions of the NGC 7538 complex, 
which contains the majority of red
sources but, in contrast, a far smaller number of Class II and Class I
sources. Presumably the star forming activity has already been over in this
region except for the interface with the molecular cloud to the south-west,
where we notice a few red sources.

Secondly, a distinct core of star forming activity is found near the
center of our survey area. It is a compact region around IRS 1-3 surrounded
by the bright IR nebula. These IR sources are all newly formed OB-star 
candidates. Together with the (ultra-) compact H II regions
which they have started to develop, they are deeply embedded in the dense
molecular core. Here we also notice a concentration of red sources. Most
probably extensive star forming activity is currently taking place in this
region. 

Turning our eyes further to the south-east, we find a rather scattered
distribution of red sources and Class II candidates. We consider they
comprise the third region of star forming activity in the NGC 7538 complex.
It seems to be composed of two substructures; one corresponds to the fluffy
reflection nebula associated with IRS 9, 9N1, 9N2 and 9N3 (the south-eastern
region), the other is the region surrounding IRS 11 and an elongated
grouping of red stars (the southern region). The extensive reflection nebula 
is a manifestation of massive outflows associated with IRS 9 
(Tamura et al. 1991). 
The presence of massive outflows associated with IRS 9 and the existence of
masers and NGC 7538 S (Class 0 candidate, \S 3.8.3) around IRS 11 are a
clear signpost of an early phase of star formation in these regions. Therefore,
this third region, both south-eastern and southern, is located in the
molecular cloud and must be very young in terms of star formation. 

The question can then be raised: what is the time sequence of
star formation history in NGC 7538 ? The central region appears to be
the result of the propagation of star formation activity from the
north-west region due to the expansion of the H II region and the
compression of the molecular cloud in the north-western interface.
Its location adjacent to the NGC 7538 H II region and the dense
concentration of YSOs between H II region and IRS 1-3 core match
quite well the characteristics of sequential star formation
(Elmegreen \& Lada 1977). The complex to the south and south-east might
have evolved independently from the same molecular cloud. Since this
region is separated from the optical H II region of NGC 7538, the star
formation here is independent of the action of its expansion, and appears to
have been taking place spontaneously, as suggested by
McCaughrean, Rayner \& Zinnecker (1991).
Alternatively, it could be possible that some larger scale
trigger might have played a role, resulting in the entire sequence of
these three stages of star formation activity. In order to answer
this question, however, it might be crucial to reveal 
the YSO population on a much larger scale.

\subsection{Individual sources and regions}

In Fig. 12 we present some selected areas of the NGC 7538 star forming region
in our high resolution (0\arcsec.7) NIR images that are of particular interest.

\subsubsection{IRS 1, 2, and 3}

NGC 7538 IRS 1--3 is a group of three infrared sources in the very bright red 
core in the center of the image (Fig. 12a). These three high luminosity 
infrared sources are located at the boundary of the H II region 
(see Fig. 1). They were found in a 2.2 $\mu$m and 
20 $\mu$m survey of NGC 7538 (Wynn-Williams et al. 1974), south-east of the 
optical H II region.
IRS 1 is identified as an ultracompact (UC) H II region, while IRS 2 and 3 are 
compact and optically thin compact H II regions, respectively (Campbell 1984; 
Campbell \& Persson 1988).
IRS 2 is embedded in the center of the compact IR nebula, whereas 
IRS 1 and IRS 3 are located near the south and south-western borders of the 
nebula, respectively (Fig. 12a). 

IRS 1 is the most luminous (L$_{NIR}$ = 1.5$\times$10$^5$ L$_{\odot}$;
Willner 1976; Hackwell et al. 1982) 
of the three infrared sources. This source is detected only in H and K$_s$-band 
among our images (Table 1). The high spatial resolution submillimeter data 
resolved the young UC H II region around IRS 1 and it was shown that the
UC H II region is  
surrounded by a cluster of submillimeter sources, none of which have near or 
mid-infrared counterparts (Sandell \& Sievers 2004). We see an elongated
group of red stars in a filamentary structure around IRS 1 in our NIR image
(see Fig. 8). IRS 2 inferred to be an O9.5 ZAMS star with a luminosity of 
$\sim$ 3.8$\times$10$^4$ L$_{\odot}$ (Campbell \& Persson 1988; 
Campbell \& Thompson 1984) based upon its ionizing flux, 
is situated $\sim$ 10\arcsec~north of IRS 1 and 
possesses the most extended H II region of the cluster sources. A red star
is located about 2\arcsec~north-west of IRS 2 (Fig. 8). 
IRS 3, the least luminous of the cluster sources is situated about 
15\arcsec~west of IRS 1. No red stars are seen close to IRS 3 (Fig. 8).

\subsubsection{IRS 9 reflection nebula}


Located about 2\arcmin~south-east of IRS 1, IRS 9 is a bright, pointlike
IR source at the apex of a complex reflection nebula (marked by 
``9'' in Fig. 12b). IRS 9 is detected only in our H and 
K$_s$-band images (Table 1). The CO outflow 
driven by IRS 9 (Kameya et al 1989) is associated with extensive H$_2$ 
emission (Fig. 5 and Davis et al. 1998). Sandell \& Sievers (2004) found 
IRS 9 to be
extended and disk-like in their high spatial resolution submillimeter maps.
The total mass of the IRS 9 submillimeter source is $\sim$ 150 M$_{\odot}$.
The bolometric luminosity of IRS 9 is $\sim$ 2$\times$10$^4$ L$_{\odot}$
(Werner et al. 1979; Thronson \& Harper 1979).

The striking reflection nebula in Fig. 12b is associated with IRS 9 
(Werner et al. 1979; Tamura et al. 1991), which shows very delicate
and complex features of structure and color. This suggests a complex
structure of circumstellar material and their massive YSOs.

At the easternmost tip of the nebula, four extremely red infrared
sources including IRS 9 are located (marked by the thin solid lines in 
Fig. 12b; see also Fig. 8). We designate these sources as IRS 9N1, IRS 9N2, 
and IRS 9N3 (see Figs. 1, 7 and Table 1). 
Two of them near the tip are associated with a ``silhouette shell''. 
While comparing with the polarization map (Tamura et al. 1991),
we find that IRS 9N1 is probably not an independent source
but is most likely an unresolved gaseous knot in the nebula,
because the degree of polarization is very high at the position of IRS 9N1 and
the polarization pattern is not disturbed. 
For other sources (IRS 9N2 and 9N3) the polarization is locally small or 
the pattern is disturbed.
These red sources as well as a nearby bluer source IRS N4 (marked by ``N4'' 
in Fig. 12b; see Table 1) are most probably very young stars in their earliest
evolutionary phases and appear to be associated with small scale 
circumstellar matter that obscures the background nebula
(we call them silhouette shell).
It is possible that the kinematics of the IRS 9 region is influenced by 
these YSOs (Sandell \& Sievers 2004).     

\subsubsection{IRS 11 and NGC 7538 S}

IRS 11 is situated $\sim$ 1\arcmin~south of IRS 1 (marked by an arrow in 
Fig. 12c). This IRS source is associated with a tiny nebulosity (Fig. 12c)
and a knot of 
vibrationally excited H$_{2}$ emission (Davis et al. 1998), suggesting that 
it is most likely a young star with an outflow. 
Kameya et al. (1989) also reported a CO outflow 
in this region. The CO outflow associated with IRS 11 appears poorly 
collimated, though, roughly aligned with that associated with IRS 1. 
A large number of red young stars are also seen around IRS 11 region (Fig. 8).
This fact, together with the results
of the submillimeter polarimetry by Momose et al. (2001),
indicates that the IRS 11 region is younger than the IRS 1 region. 

NGC 7538 S ($\alpha_{2000}$, $\delta_{2000}$ = $23^h13^m44^s.51$, 
+61$^{\circ}26^{\arcmin}48^{\arcsec}.7$ at 850 $\mu$m), 
a high mass Class 0 candidate, 
is about 80\arcsec~to the south of IRS 1 (850 $\mu$m position is shown by 
a cross symbol in Fig. 12c; Sandell \& Sievers 2004). 
These authors resolved the submillimeter emission into an 
elliptical source (NGC 7538 S) of 
14\arcsec $\times$ 7\arcsec~P.A. = 58$^{\circ}\pm3^{\circ}$. 
We find two very red sources associated with this elliptical source 
($\alpha_{2000}$, $\delta_{2000}$ = $23^h13^m44^s.65$, 
+61$^{\circ}26^{\arcmin}47^{\arcsec}.9$; $23^h13^m43^s.96$, 
+61$^{\circ}26^{\arcmin}44^{\arcsec}.9$; see Figs. 8 and 12c). 
However, they are unlikely to be the counterparts to the
submillimeter source, considering the Class 0 nature of NGC 7538 S. 
The existence of NGC 7538 S and a concentration of 
red sources further support the conclusion that the southern region is 
younger than other regions in NGC 7538.

\subsubsection{Bow shock region?}

Fig. 12d is a section in the north-east corner of our JHK$_s$ image in 
Fig. 3, where we detect peculiar nebulosities. The northern one shows a 
cone-shaped silhouette enveloped by a faint nebulosity, probably, due to 
bow-shaped H$_2$ emission (Fig. 5 and Davis et al. 1998). The southern 
one shows an oval-shape overlaid by dark patches,
which is associated with at least two YSOs (see Fig. 8). This nebula is 
within a local deep obscuration in an optical image.
About 40\arcsec~north of this object a bow-shaped structure of H$_2$ emission,
which is made of two elongations, one in E-W and the other in N-S, is found
(Fig. 5, see also Fig. 7 of Davis et al. 1998). This structure is presumably
caused by an outflow from the north-east of this region.

\subsubsection{NGC 7538 IRS 4}

Fig. 12e shows the barely resolved double source IRS 4
(separation $\sim$ 1\arcsec4) in our near-infrared
images (marked by an arrow in the center of the image). It is an isolated red 
source (H-K = 1.94) associated with local nebulosity, which is located 
towards the north-west of IRS 2 at the boundary of the optical H II region. 

A faint, extended dust condensation was seen about 15\arcsec~in the south-west 
of IRS 4 in 
850 $\mu$m map (Sandell \& Sievers 2004). These authors also found 
a fainter extended submillimeter source near the 20 $\mu$m source IRS 4. IRS 4,
in view of its significant 20 $\mu$m flux, is most probably a compact H II
region. It is also situated close to a knot in the ammonia emission and
appears to be close to the interface between the H II region and the molecular
cloud (Zheng et al. 2001).   
 
\section{Conclusions}

A sub-arcsec JHK$_s$-band NIR imaging survey of YSOs associated with the 
NGC 7538 star forming region is presented. The survey covers a 
4\arcmin.9$\times$4\arcmin.9 area down to a 
limiting magnitude (10 $\sigma$) of J = 19.5, H = 18.4, and K$_s$ = 17.3.
The NIR images presented in this work are deeper than any JHK surveys 
to date for the larger area of NGC 7538 star forming region.
From the analysis of these images we derive the following conclusions : 

1) We see several different evolutionary stages in the NGC 7538 star 
forming complex with considerable substructure, as also suggested by
McCaughrean, Rayner, \& Zinnecker (1991). There are lines of  
evidence for sequential star formation in NGC 7538.

2) Most of the YSOs in the NGC 7538 are arranged in the north-west and 
south-east regions, which form a sequence in age: the diffuse H II region 
(north-west, oldest : where most of the Class II and Class I sources are 
detected); the compact IR core (center); and the 
regions with the extensive IR reflection 
nebula and a cluster of red young stars (south-east and south).

3) A large number of red stars (H-K $>$ 2) are detected in the NGC 7538 
molecular cloud region, most of which are clustered around the molecular 
clumps associated with IRS 1--3 and toward south and south-east of IRS 1--3. 

4) The YSOs in the central region are probably the results of the
propagation of star forming activity from the north-western region due to
the expansion of the H II region and the compression of the molecular cloud
(sequential star formation, Elmegreen \& Lada 1977). The
south-eastern/southern region is independent of the above action and
presumably the star formation there is taking place in a spontaneous and
gradual process.

5) The KLF of the whole NGC 7538 region shows a power-law slope : 
$\alpha$ = 0.30$\pm$0.03, which is lower than the typical values reported
for the embedded young clusters, although equally low values have also
been reported in the W3 Main star forming region. We also find a small 
increase in the KLF slope from the ``younger'' region ($\alpha$ $\sim$ 0.27) 
to the most evolved ``older'' region ($\alpha$ $\sim$ 0.33) in NGC 7538. 
The resulting KLF slopes may therefore indicate the age sequence of various 
regions in NGC 7538.  

6) Using the age of $\sim$ 1 Myr, we find that 
about 88\% of the YSO candidates have an upper mass limit of 3 M$_{\odot}$. 
We estimate that the lowest mass limit of Class II and Class I candidates 
in our observations is 0.1 M$_{\odot}$.
Therefore, the stellar population in NGC 7538 is primarily
composed of low mass PMS stars. 

7) The radio continuum image based on the GMRT
observations at 1280 MHz shows interesting morphological details,
including an arc-shaped structure highlighting the interaction
between the H II region and the adjacent molecular cloud. The ionization
front at the interface between the H II region and the molecular cloud is
clearly seen by comparing the radio continuum, molecular line, and molecular
hydrogen images. It is most plausible that NGC 7538 IRS 6 is the main 
exciting source responsible for the H II region.   

\acknowledgments

It is a pleasure to thank the anonymous referee for a most thorough reading
of this paper and several useful comments and 
suggestions, which greatly improved the scientific content of the paper.
We thank the staff of the University of Hawaii 2.2m telescope for supporting 
the first run of SIRIUS. D.K.O. was supported by the Japan Society for the 
Promotion of Science through a fellowship, during which most of this work was 
done. We acknowledge 
support by Grant-in-Aid (10147207, 12309010, 13573001) from the Ministry of 
Education, Culture, Sports, Science, and Technology.  
We thank the staff of the GMRT that made the radio observations possible.
The GMRT is run by the National Centre for Radio Astrophysics of the Tata
Institute of Fundamental Research (India).
We thank Francesco Palla for providing us with the PMS grids.
We thank Annie Robin for letting us use their model of stellar population
synthesis. We thank Chris Davis for providing us with the FITS images of
his H$_2$, narrow-band K continuum and CO molecular line of the NGC 7538. 
This publication makes use of data products from the Two Micron All Sky
Survey, which is a joint project of the University of Massachusetts and the
Infrared Processing and Analysis Center/California Institute of Technology,
funded by the National Aeronautics and Space Administration and the National
Science Foundation.

\clearpage

\begin{table}
\caption{Bright infrared sources associated with the H II region and molecular 
clump (labeled as IRS numbers in Fig. 7}
\footnotesize
\begin{tabular}{ccccccc} 
\\
\tableline
\tableline
Source  & RA (2000)   & DEC (2000)  & J     &   H   & K & Remark/Sp. Type\\
        & $hh~mm~ss$  & $dd~mm~ss$  & mag   & mag   & mag   &  (From CM diagram)\\
\tableline
IRS 1   & 23 13 45.32 & +61 28 11.7 &       & $^{**}$14.10$\pm$0.10 &  $^{**}$8.90$\pm$0.10 & \\
IRS 2   & 23 13 45.47 & +61 28 19.9 & $^{*}$11.83$\pm$0.06 & $^{*}$10.33$\pm$0.08 &  $^{*}$9.18$\pm$0.10 & O5\\ 
IRS 3   & 23 13 43.65 & +61 28 14.0 & 16.56$\pm$0.04 & 13.63$\pm$0.03 & $^{**}$11.60$\pm$0.10 & O6-O9\\
IRS 4  & 23 13 32.39 & +61 29 06.2 &  & $^{*}$9.74$\pm$0.05 & $^{*}$7.80$\pm$0.01 & \\
IRS 5   & 23 13 30.24 & +61 30 10.3 & $^{*}$10.38$\pm$0.03 &  $^{*}$9.82$\pm$0.04 &  $^{*}$9.43$\pm$0.04 & O9\\
IRS 6   & 23 13 34.38 & +61 30 14.6 &  $^{*}$8.71$\pm$0.02 &  $^{*}$8.21$\pm$0.04 &  $^{*}$8.00$\pm$0.01 & O5-O6\\
IRS 7   & 23 13 36.80 & +61 30 39.6 &  $^{*}$8.44$\pm$0.04 &  $^{*}$7.63$\pm$0.04 &  $^{*}$7.37$\pm$0.01 & \\
IRS 8   & 23 13 27.58 & +61 30 50.1 & 12.09$\pm$0.02 &  $^{*}$9.94$\pm$0.03 &  $^{*}$9.03$\pm$0.02 & O5\\ 
IRS 9   & 23 14 01.76 & +61 27 19.9 &       & 15.70$\pm$0.04 & $^{*}$10.19$\pm$0.05 & \\
IRS 9N1  & 23 14 01.36 & +61 27 18.4 & 19.79$\pm$0.11 & 15.46$\pm$0.08 & $^{*}$11.32$\pm$0.08 & \\
IRS 9N2  & 23 14 02.70 & +61 27 14.1 &                & 18.39$\pm$0.08 & 15.89$\pm$0.03&\\
IRS 9N3  & 23 14 02.56 & +61 27 24.6 &                & 17.72$\pm$0.05 & 15.22$\pm$0.07&\\
IRS 9N4  & 23 14 02.40 & +61 27 20.4 &                & 17.84$\pm$0.11 & 15.39$\pm$0.10&\\
IRS 11  & 23 13 43.92 & +61 26 57.8 & 18.50$\pm$0.03 & 15.02$\pm$0.03 & 12.88$\pm$0.04 & B0\\
\tableline
\end{tabular}
$^*$ 2MASS magnitudes.

$^{**}$ Magnitudes are from Bloomer et al. (1998). 
\end{table}

\clearpage

\begin{table}
\caption{Power-law fits to KLFs in NGC 7538}
\begin{tabular}{ccc} 
\\
\tableline
\tableline
Region  & $\alpha$ & \\
\tableline
Whole NGC 7538                & 0.28$\pm$0.02 & completeness corrected \\  
                              & 0.30$\pm$0.03 & completeness corrected, foreground,\\
                              &               & and background subtracted \\
\hline
Younger (IRS 1--3, 9, and 11) & 0.27$\pm$0.03 &\\
\hline
Older (optical H II)          & 0.33$\pm$0.04 &\\
\hline
\tableline
\end{tabular}
\end{table}

\clearpage

\begin{figure}
\epsscale{0.8}
\caption{J, H, and K$_s$-band images of the NGC 7538 star forming region 
displayed in a logarithmic intensity scale. The field of view is 
$\sim$ 4\arcmin.9$\times$4\arcmin.9. The locations of the individual
IR sources are marked in the K$_s$-band image. The direction of the NGC 7538 
IRS 9 ``H$_2$ flow'' reported by Davis et al. (1998) is also indicated in the 
K$_s$-band image. North is up and east is to the 
left. The abscissa and the ordinate are in J2000 epoch.
\label{fig1}}
\end{figure}

\clearpage

\begin{figure}
\epsscale{0.8}
\caption{GMRT low resolution map of the whole NGC 7538 region at 1280 MHz. The 
resolution is 22\arcsec.5$\times$12\arcsec.4 along PA = 4.6$^\circ$, 
and the rms 
noise in the map is $\sim 4$ mJy beam$^{-1}$. The abscissa and the ordinate 
are in J2000 epoch. The inset is a GMRT high resolution map of the central 
radio peak. The high resolution image has a resolution of 
6\arcsec.6 $\times$ 4\arcsec.7 along PA = 21.4$^\circ$, an rms of 
$\sim 1.3$ mJy beam$^{-1}$, and a total flux density of 978.7 mJy. The lower
part of the map is the same area as the JHK$_s$ image (see Fig. 4).
\label{fig2}} 
\end{figure}

\clearpage 

\begin{figure}
\epsscale{1}
\caption{JHK$_s$ composite image of the NGC 7538 star forming 
region (J: blue, H: green, K$_s$: red) obtained with the three-color SIRIUS 
infrared array mounted on the UH 2.2 m telescope. The field of view is 
$\sim$ 4\arcmin.9$\times$4\arcmin.9. North is up and east is to the left.
\label{fig3}} 
\end{figure}

\clearpage

\begin{figure}
\epsscale{0.9}
\caption{SIRIUS K$_s$-band image overlayed by the GMRT radio continuum 
contours in blue color. The red contours show the integrated emission of the 
CO (J = 2--1) line (from Davis et al. 1998). The CO contours start at 
50 K km s$^{-1}$ and increase in steps of 15 K km s$^{-1}$. The CO data are
taken with JCMT at $\sim$ 20\arcsec~resolution. The positions of the IRS 
sources (see Table 1) are marked by crosses. 
The abscissa and the ordinate are in J2000 epoch.
\label{fig4}}
\end{figure}

\clearpage

\begin{figure}
\caption{Continuum-subtracted molecular hydrogen emission at 2.122 $\mu$m
of the NGC 7538 region (Davis et al. 1998). Residual artifacts left over
by the subtraction of the 2.140 $\mu$m image from the 2.122 $\mu$m image
are seen around the positions of some bright stars. The overlaying contours
and symbols are same as shown in Fig. 4. The direction of the IRS 9 ``H$_2$ jet''
in the south-eastern corner
and the bow shock region in the north-eastern corner of the image
are indicated. The abscissa and the ordinate are in J2000 epoch.
\label{fig5}}
\end{figure}

\clearpage

\begin{figure}
\epsscale{1}
\caption{Color-color diagrams of (a) the NGC 7538 star forming region and 
(b) the reference field for the sources
detected in JHK$_s$-bands with photometric errors less than 0.1 mag. 
The sequences for 
field dwarfs (solid curve) and giants (thick dashed curve) are from 
Bessell \& Brett (1988). The dotted line represents the locus of T-Tauri stars 
(Meyer et al. 1997). Dashed straight lines represent the reddening vectors 
(Rieke \& Lebofsky 1985). The crosses on the dashed lines are separated by 
A$_V$ = 5 mag. 
\label{fig6}}
\end{figure}

\clearpage

\begin{figure}
\caption{H-K/K color-magnitude diagram for the sources detected in 
H and K$_s$-bands with photometric errors less than 0.1 mag. 
Stars and filled triangles represent the YSOs identified from
the regions ``T'' and ``P'' in Fig. 6a, respectively.  
The vertical solid lines from left to right indicate the  
main sequence track at 2.8 kpc reddened by A$_V$ = 0, 20, 40, and 60 
magnitudes, respectively. The intrinsic colors are taken from 
Koornneef (1983). Slanting
horizontal lines identify the reddening vectors (Rieke \& Lebofsky 1985).
Also shown are the positions of known IRS sources as open squares. 
\label{fig7}}
\end{figure}

\clearpage

\begin{figure}
\caption{Spatial distribution of the YSO color-excess candidates superposed on 
the K$_s$-band image with a logarithmic intensity scale. Blue asterisks 
represent T-Tauri and related sources (Class II), green filled triangles 
indicate Class I sources, and red filled circles denote the 
red sources (H-K $>$ 2). The abscissa and the ordinate are in J2000 epoch.
\label{fig8}}
\end{figure}

\clearpage

\begin{figure}
\caption{Comparison of the observed KLF in the reference field and the 
simulated KLF from star count modelling.
The filled circles and solid line denote the K star counts
in the reference field and the filled triangles and dotted line represent
the simulation from the galactic model (see the text). The KLF slope 
($\alpha$, see \S 3.5) of the dotted line (model simulation) is 0.29$\pm$0.01.
\label{fig9}}
\end{figure}

\clearpage

\begin{figure}
\caption{(a) and (b) Corrected K$_s$-band luminosity function for the 
whole NGC 7538 region. In (a), the dotted line denotes the raw KLF and the solid 
line shows the KLF corrected only for the completeness. In (b), 
the dotted line shows the KLF corrected only for the completeness.
The dashed line denotes the field star counts from the reference field 
corrected for the completeness and modified to reflect the unextincted 
foreground stars and the background stars reddended by A$_V$ = 30 mag 
(A$_K$ = 3.36 mag), with the help of the galactic model (Robin et al. 2003).
The solid line corresponds to the field star subtracted KLF. (c) and (d) are 
the completeness-corrected and field star-subtracted KLF of the 
NGC 7538 region, respectively. 
The solid lines are the best linear fit to the data points.
\label{fig10}}
\end{figure}

\clearpage  

\begin{figure}
\caption{Color-magnitude diagram for the YSO candidates in NGC 7538. 
Class II candidates are indicated by stars, filled triangles represent Class I 
candidates, and the filled circles are red sources with H-K $>$ 2 with 
J-band counterparts. The solid curve denotes the loci of 10$^6$ yr old PMS stars, 
and the dotted curve is for those that are 0.3$\times10^{6}$ yr old, both derived 
from the model of Palla \& Stahler (1999). Masses range from 0.1 to 
4 M$_{\odot}$ from bottom to top, for both curves. The dotted oblique reddening
line denotes the position of a PMS star with 3 M$_{\odot}$ for 0.3 Myr, and the 
solid oblique lines denote the positions of PMS stars with 0.1, 2, and
3 M$_{\odot}$ for 1 Myr, respectively, in this diagram. Most of the objects
well above the PMS tracks are luminous and massive ZAMS stars (see Table 1,
Fig. 7).
\label{fig11}}
\end{figure}   

\clearpage

\begin{figure}
\caption{Enlarged view of the color image of selected areas (see Fig. 3 and
\S 3.8). a) NGC 7538 IRS 1--3 region. The three high luminosity infrared
sources IRS 1--3 are located at the boundary of the H II region.
b) The striking reflection nebula in the south-eastern corner of our JHK$_s$ 
image in Fig. 3. NGC 7538 IRS 9 is a bright and 
pointlike infrared source at the apex of the complex reflection nebula 
(marked by ``9'').  
In its immediate vicinity three more extremely red infrared sources, 
IRS 9N1, IRS 9N2, and IRS 9N3 are located
(marked by ``N1'', ``N2'', and ``N3''; see \S 3.8.2). 
c) NGC 7538 IRS 11 (marked by an arrow) 
is most likely a young star associated with nebulosity. 
The cross shows 
the 850 $\mu$m position of NGC 7538 S (a high mass Class 0 candidate). 
d) The small peculiar nebulosities seen in the north-east 
corner of our JHK$_s$ image in Fig. 3. The northern one shows a cone-shaped 
silhouette enveloped by 
a faint nebulosity, probably due to bow-shaped H$_2$ emission 
(see Fig. 5 and Davis et al. 1998). 
The southern one shows an oval-shaped overlaid 
by dark patches. 
e) The barely resolved double source NGC 7538 IRS 4 (arrowed) associated with
the local nebulosity,
which is located towards north-west of IRS 2 at the boundary of the
optical H II region.   
\label{fig12}}
\end{figure}

\end{document}